\documentclass[useAMS,usenatbib]{mn2e}

\usepackage{amsmath}
\usepackage{comment}
\usepackage{graphicx}
\usepackage{rotating}
\usepackage{color}
\usepackage{aas_macros}
\usepackage{amsfonts}
\newcommand{\beq}{\begin{equation}}
\newcommand{\enq}{\end{equation}}
\newcommand{\beqa}{\begin{eqnarray}}
\newcommand{\enqa}{\end{eqnarray}}
\newcommand{\beit}{\begin{itemize}}
\newcommand{\enit}{\end{itemize}}
\newcommand{\bem}{\begin{pmatrix}}
\newcommand{\enm}{\end{pmatrix}}

\newcommand{\vecC}{\mathbf C}

\newcommand{\vecx}{\mathbf{x} }
\newcommand{\vecy}{\mathbf{y} }
\newcommand{\vecz}{\mathbf{z} }

\newcommand{\vectheta}{\bm{\theta}}
\newcommand{\veck}{\mathbf{k} }

\newcommand{\vecq}{\mathbf{q}}

\newcommand{\lat}{\left\langle}
\newcommand{\rat}{\right\rangle}
\newcommand{\av}[1]{\lat #1 \rat}

\newcommand{\lb}{\left [}
\newcommand{\rb}{\right ]}
\newcommand{\lp}{\left (}
\newcommand{\rp}{\right )}

\newcommand{\bes}{\begin{sideways}}
\newcommand{\ees}{\end{sideways}}

\renewcommand{\bbeta}{\boldsymbol {\eta}}
\renewcommand{\bmu}{\boldsymbol {\mu}}
\newcommand{\vecl}{\mathbf l}

\newcommand{\vecn}{\textbf{n}}

\newcommand{\vecN}{\mathbf N}
\newcommand{\vecrho}{\boldsymbol \rho}
\newcommand{\vecA}{\mathbf A}

\newcommand{\vecf}{\mathbf f}
\newcommand{\vecepsilon}{\boldsymbol \epsilon}
\renewcommand{\vectheta}{\boldsymbol \theta}

\title[Generating Cosmological Random Fields]{On fast Generation of Cosmological Random Fields}
\author[Carron, Wolk and Szapudi]{J. Carron\thanks{E-mail:
carron@ifa.hawaii.edu}, M. Wolk and I. Szapudi  \\
Institute for Astronomy, University of Hawaii, 2680 Woodlawn Drive, Honolulu, HI, 96822}
\begin{document}

\date{\today}

\pagerange{\pageref{firstpage}--\pageref{lastpage}} \pubyear{2014}

\maketitle

\label{firstpage}

\begin{abstract}
The statistical translation invariance of cosmological random fields is broken by a finite survey boundary, correlating the observable Fourier modes. Standard methods for generating Gaussian fields either neglect these correlations, or are costly, or both. Here we report on a fast and exact simulation method applicable to a wide class of two-point statistics that requires the simulation of a periodic grid of only twice the survey side with fast Fourier transforms. Super-survey modes, dominating the covariance of power spectra beyond linear scales in galaxy surveys and causing the correlation of large and small scales, ``beat coupling'',  or "super-sample" covariance, are precisely accounted for in non-linear transformations of the Gaussian field. As an application, we simulate the CFHTLS  $\sim 7^\circ  \times 7^\circ$ W1 galaxy density field, modeled as a Poisson sampling of a lognormal density field. We show that our simulations produce power spectra, $A^*$-power spectra, counts-in-cells probability distributions as well as covariances perfectly consistent with the data. In addition, our technique reproduces the information plateau beyond linear scales as observed previously in Sloan Digital Sky Survey galaxy catalogs and in $N$-body simulations. Our method is thus an efficient yet powerful simulation and prediction tool for galaxy survey data and covariances.

\end{abstract}

\section{Introduction and overview}
\indent
In the widely accepted inflationary paradigm, cosmological fields such as the matter density are represented as homogeneous (and isotropic) random fields. Such fields have translation invariant statistics, allowing a convenient description in terms of uncorrelated Fourier modes. 
Thus an instance of a Gaussian field, such as the initial state of the matter field in an $N$-body simulation, is conceptually easy to generate. The uncorrelated Gaussian modes can be generated from their variance, the power spectrum, and the field obtained by Fourier transformation to real space, usually with efficient Fast Fourier Transform (FFT) algorithms  \citep{PressEtal07}.\newline
\indent
Consider now the task of generating a homogeneous Gaussian field on a regular grid (e.g. 'pixels' or 'cells'), not for a $N$-body simulation, but in order to simulate directly cosmological random fields, such as galaxy surveys. For these, statistical independence of the observed Fourier modes is no longer true: the finite survey volume breaks the translational symmetry of the field, thus correlating the observable Fourier modes. Our goal is to discuss a fast and exact method for this case, where the field values on the $N_{\textrm{cells}}$ cells obey the target $N_{\textrm{cells}}$-variate Gaussian probability density function (PDF), to numerical precision. This means that the $N_{\textrm{cells}} \times N_{\textrm{cells}}$ covariance matrix of the Gaussian variables must match a non-diagonal target matrix. Covariance matrix factorization techniques can always produce exact simulations, but their cost scales at best with $N^2_{\textrm{cells}}$. They are thus often costly and impractical, especially when a large number of simulations is desired. The standard procedure is to enforce translation invariance by imposing periodic boundary conditions outside the desired volume, or a larger volume simulated with the use of FFT algorithms. This suppresses the modes larger than the simulated volume, making the simulation only approximate, with errors that are not always simple to evaluate. The importance of such 'super-survey' modes in the context of galaxy surveys have been stressed in many theoretical studies, starting from \citet{RimesHamilton06} and \citet{HamiltonEtal06}. They are responsible for most of the covariance matrix of the power spectrum beyond linear scales, due to 'beat-coupling'  \citep{RimesHamilton05,RimesHamilton06,NeyrinckEtal06,SefusattiEtal06,TakahashiEtal09,SchneiderEtal11,DePutterEtal12,TakadaHu13,LiEtal14}, causing a saturation in the information content of the power spectrum on scales larger than naively expected. This saturation was also observed in the SDSS data galaxy angular power spectrum by \citet{LeePen08}.
\newline
\indent
Our principal aim is to present an exact simulation technique, based on circulant embedding and applicable to a wide class of homogeneous two-point functions. It requires only the simulation of a grid of twice the survey size with FFT methods. It is thus fast and convenient despite incorporating super-survey modes exactly. To best of our knowledge absent from the cosmological literature, the method was described in a journal of hydrological sciences by \citet{DietrichNewsam93} (see also \citet{WoodChan94,DietrichNewsam97}). The method shares some similarities with the '$\xi$-sampling' method of \citet{Pen97} and \citet{Sirko05} for the generation of initial conditions (IC) for $N$-body simulations. There, the power spectrum is first convolved with a window function prior to sampling the modes, in order to better reproduce the real space statistical properties of the field. This stands in contrast to the '$P$-sampling', or $k$-space sampling method \citep[e.g.,][]{Bertschinger01} for IC generation, where the discrete modes are simply set to match that of the continuous field to be simulated, and thus do not contain super-survey modes \citep{Sirko05,GnedinEtal11}.
\newline
\indent
As an application of the circulant embedding technique we simulate galaxy count maps in square cells of the projected Canada-France-Hawaii Telescope Large Survey (CFHTLS\footnote{\texttt{http://www.cfht.hawaii.edu/Science/CFHLS/}})  in the flat-sky approximation. We model the galaxy counts Poisson-sampling an underlying, lognormal density field \citep{ColesJones91}. The lognormal field is simple and reproduces qualitatively the long tail in the matter field PDF \citep{KofmanEtal94,BernardeauKofman95,KayoEtal01}. The latter causes the spectrum and higher order statistics to capture information inefficiently \citep{Carron11,CarronNeyrinck12}. Lognormal statistics are often used in galaxy surveys to produce estimates of covariance matrices \citep{BlakeEtal11,ColeEtal05,BeutlerEtal11}. We show that for the particular case of the CFHTLS our simulation quantitatively reproduces both the measured power spectra and higher order statistics.
\newline
\indent
This paper is structured as follows. In section \ref{circulantembedding} we present the circulant embedding method. We begin with the case of a continuous field in a finite volume then show that the method remains exact on a discrete grid, when implemented as discussed at the end of the section. In section \ref{application}, we first discuss our model for the CFHTLS count maps. We then compare various model predictions to the measurements by direct comparison to a large number of simulations. We discuss the power spectrum, count in cells PDF, and the spectrum of a non-linear transformation of the data map.  We also illustrate, with the help of the information plateau in the power spectrum discussed above, the importance of including the super-surveys modes by looking at the corresponding outputs from simulations with periodic boundary conditions. We conclude in section \ref{Conclusions}. An appendix presents more details about the implementation of our model for the count maps.


\section{The Circulant embedding method for fast generation of Gaussian fields}
\label{circulantembedding}

\begin{figure}
\begin{center}
\includegraphics[width = 0.45\textwidth]{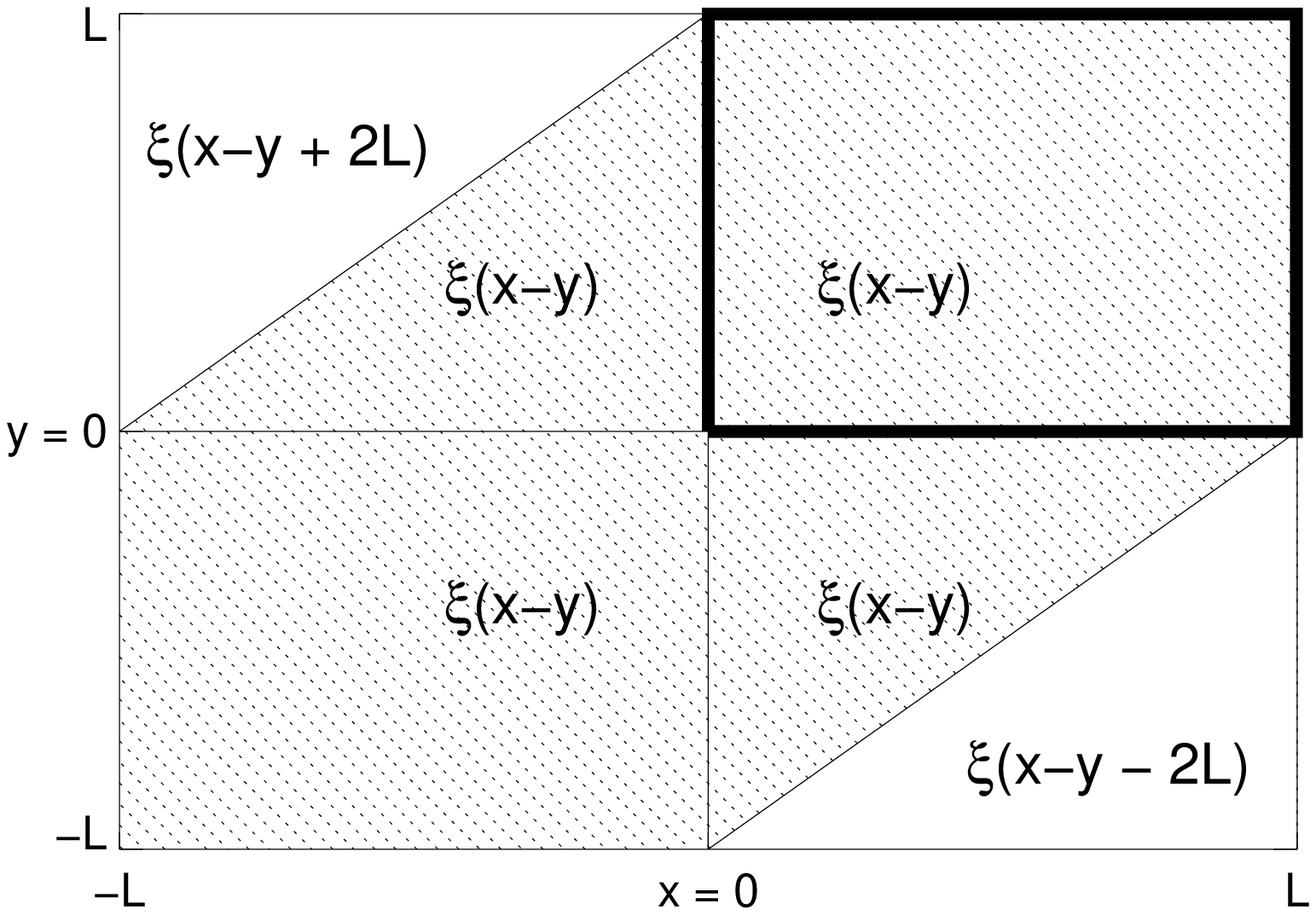}
\caption{A schematic representation of the embedding of the covariance matrix in the circulant embedding method (here for a one-dimensional field for clarity), in order to simulate a Gaussian field in a volume $[0,L]$ with two-point function $\xi(x-y)$, with no periodicity at the boundaries. The volume is extended to $x$ in $[-L,L]$, with covariance function $\xi(x,y)$ between the new field values assigned as illustrated in the figure. The upper right square corresponds to the covariance matrix of the original volume. In the dashed region, the covariance is set to the two-point function $\xi(x-y)$. The lower right and upper left corners are set to the unphysical values $\xi(x-y - 2L)$ and $\xi(x-y + 2L)$ respectively. In this way, the larger covariance matrix is now exactly diagonalized by the Fourier transform (the covariance matrix is now a 'circulant'). Any subset of length $L$ of the larger volume simulated with independent Fourier modes has covariance function in the shaded area. Thus, it possesses the desired covariance $\xi(x-y)$, and the simulation is exact. It is key that the method remains exact on a finite grid. The method only works if the enlarged covariance function is still positive definite, i.e. if the left hand side of Eq. \eqref{Specd} is positive.}
\label{figcirculantembedding}
\end{center}
\end{figure}
\indent
The idea is simpler to grasp and to illustrate in the continuous limit of the method, which we discuss first. We are interested in generating a $n$-dimensional Gaussian field in a hypercube $[0,L]^n$, with exact covariance function $\xi(\vecx-\vecy),$ with $\vecx, \vecy \in [0,L]^n$ (the implementation for hyperrectangles only requires trivial modifications). 
Extending $[0,L]$ to $[-L,L]$, we define a power spectrum through
\beq \label{Padef}
\tilde P(\veck) = \int_{[-L,L]^n} d^nx\: \xi(\vecx)\:e^{-i\veck \cdot \vecx}.
\enq
If it is found that
\beq \label{condition}
\tilde P(\veck) \ge 0
\enq
for all discrete Fourier modes $\veck$ of the larger box, we can then generate an homogeneous Gaussian random field $\phi(\vecx)$ with periodic boundary conditions on $[-L,L]^n$ requiring independent, zero mean Gaussian Fourier modes:
\beq \label{Papos2}
\av{\tilde \phi(\veck) \tilde \phi^{*}(\vecq)} =  \delta_{\veck\vecq}  \:\tilde V\tilde P(\veck),\quad \tilde V =(2L)^n.
\enq
Note that the method fails if condition \eqref{condition} is not met, for Eq. \eqref{Papos2} requires $\tilde P(\veck)$ to be positive.
The restriction of $\phi(\vecx)$ to the desired volume $[0,L]^n$ has precisely the two-point function searched for.
\beq \label{thatiswhatwewant}
\av{\phi(\vecx) \phi(\vecy)} = \xi(\vecx-\vecy), \textrm{    for  }\vecx,\vecy \in [0,L]^n. 
\enq
To see this, explicit calculation of the covariance with definitions \eqref{Padef} and \eqref{Papos2} gives
\beq
\av{\phi(\vecx) \phi(\vecy)} = \frac 1 {\tilde V} \int_{[-L,L]^n} d^nz \:\xi(\vecz) \lp \sum_{\veck} e^{i \veck \cdot \lp \vecx - \vecy -\vecz \rp}\rp.
\enq
The sum on the right is the Dirac-comb
\beq
\tilde V \sum_{\vecn \in \mathbb Z^n} \delta^D\lp \vecx - \vecy -\vecz + 2L\vecn\rp.
\enq
If both $\vecx$ and $\vecy$ are restricted to the sub-volume $[0,L]^n$ (or any sub-volume of the form $[a,a+L]^n$), the contributing term is $\vecz = \vecx - \vecy$ and we recover Eq. \eqref{thatiswhatwewant} as claimed. For a pair of points $\vecx$ and $\vecy$ outside our subset, the covariance between $\phi(\vecx)$ and $\phi(\vecy)$ is $\xi$ at (unphysical) arguments $\vecx -\vecy \pm 2L$. This is illustrated in Fig. \ref{figcirculantembedding}.
\newline
\newline
\indent
A slight modification of the above algorithm will provide exact simulations for a finite number of grid points.  Given the desired set of grid points (to simplify notation, we assume a two dimensional field from here on)
\beq
 \vecx  = \lp\frac{L}{N} \rp \:\begin{pmatrix} i \\ j \end{pmatrix},\quad i,j =0,\cdots,N-1,
\enq
we enlarge it to the set
 \beq \label{grid}
 \vecx  = \lp\frac{L}{N} \rp \:\begin{pmatrix} i \\ j \end{pmatrix},\quad i,j = -N,\cdots,N-1.
 \enq
 Given the input two-point function $\xi$, the 'ad-hoc' power spectrum is defined as the discrete Fourier transform
 \beq \label{Specd}
 \tilde P(\veck) = \lp\frac{L}{N} \rp^2 \sum_{\vecx \textrm{ in Eq. }\eqref{grid}}\xi(\vecx) e^{-i\veck\cdot \vecx},
 \enq
 for the corresponding modes
 \beq
 \veck = \lp \frac{2\pi}{2L} \rp \begin{pmatrix} n_1 \\ n_2 \end{pmatrix},\quad n_1,n_2 = 0,\cdots,2N-1.
 \enq
 The spectrum is trivially evaluated with FFT.  The discrete Fourier modes of the field on the grid are defined as in Eq. \eqref{Papos2}, with the spectrum in Eq. \eqref{Specd}. The map is obtained from these modes by the inverse discrete Fourier transform. The covariance between the field values at two grid points $\vecx$ and $\vecy$ inside the desired sub-volume remains $\xi(\vecx-\vecy)$ despite the discreteness of the transforms and of the grid.
 This can be shown by a very similar calculation than above, the Dirac-comb replaced by a Kronecker delta. Thus, the discrete simulation is exact. Again, the spectrum in Eq. \eqref{Specd} must be positive for the method to work. While this is only guaranteed for asymptotically large $L$, we did not encounter a single problematic case in the application to which we now turn to.
 \section{Application to galaxy fields in the CFHTLS} \label{application}
\begin{figure*}
\begin{center}
\includegraphics[width = 0.45\textwidth]{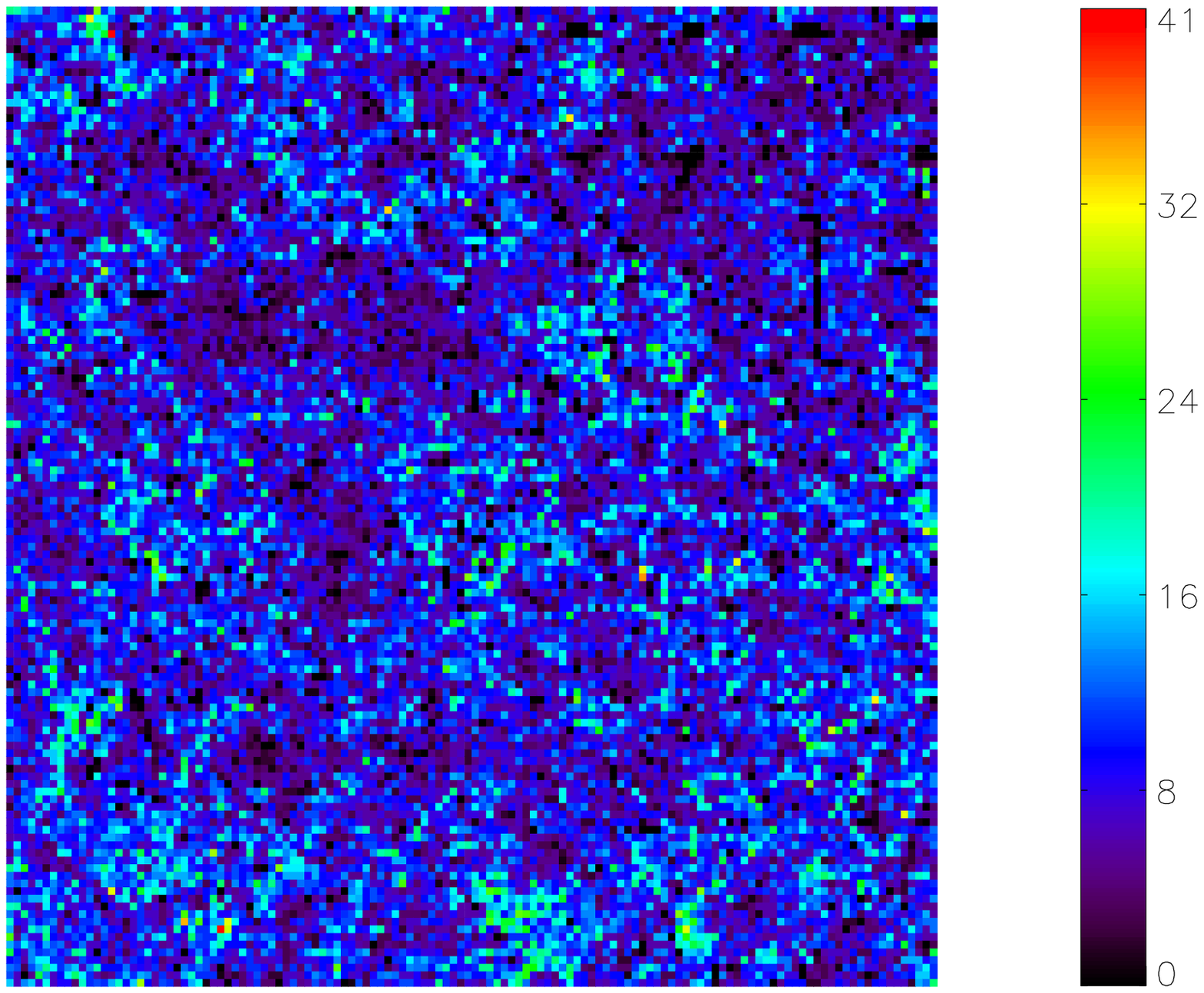}
\includegraphics[width = 0.45\textwidth]{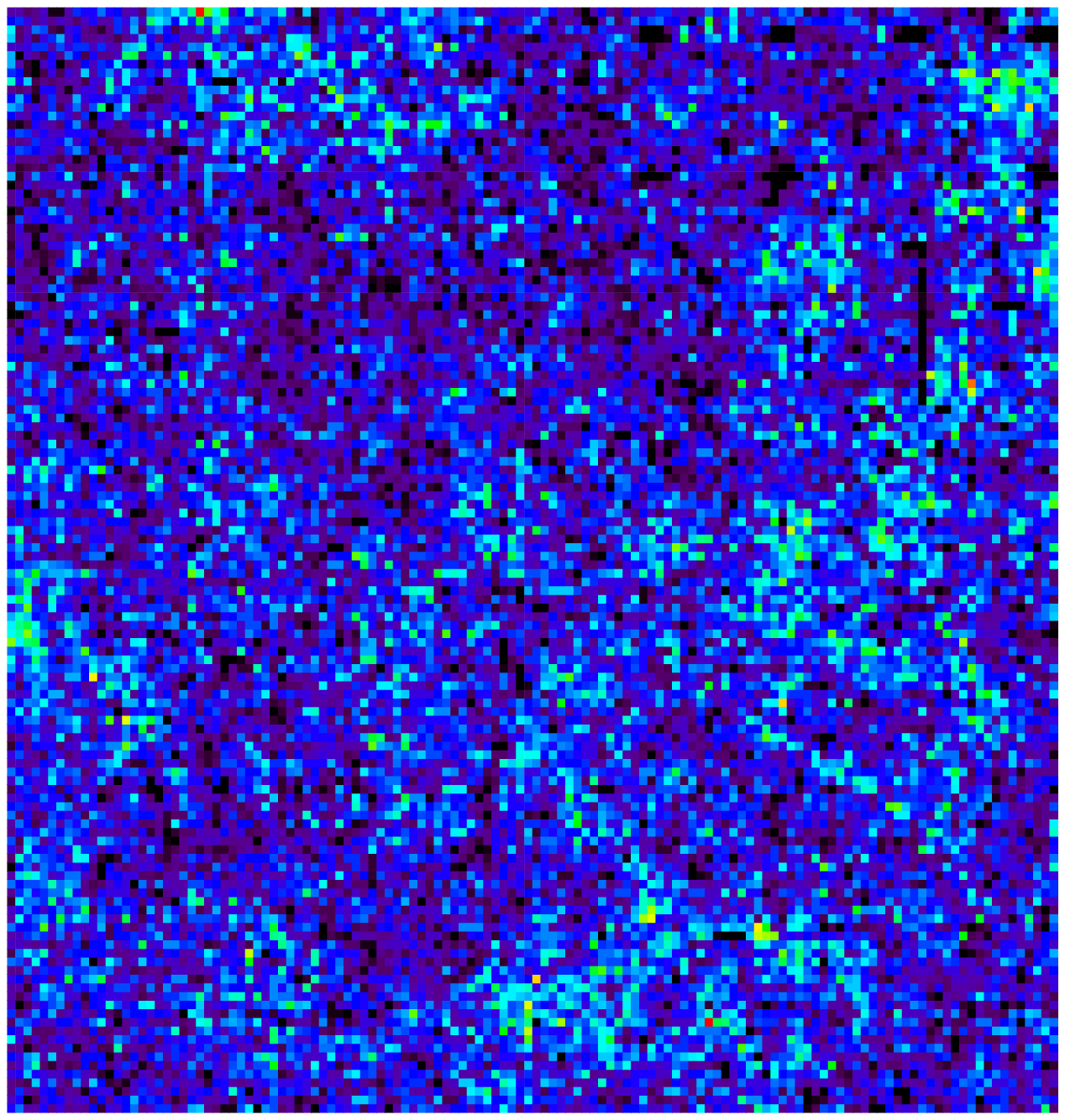}
\caption{\label{Figmaps} The left panel shows the galaxy counts in $128^2$ cells in the $7.46 \times 7.46$ deg$^{2}$ CFHTLS W1 field, including galaxies with photometric redshift estimate between 0.6 and 0.8. The right panel shows a Poisson sampling of a $128^2$-variate lognormal density field, generated with the circulant embedding method exposed in the text. The unfiltered two-point function of that underlying field matches the best-fit model from the HOD analysis of \citet{WolkEtal13}. The variance in cells of the underlying density fluctuations in the right panel is $\sigma^2_\delta = 0.14$, and the Poisson intensity $\bar N  =9.68$. The color code is identical on the two panels.}
\label{figdatasim}
\end{center}
\end{figure*}

 \indent
 We apply the above technique to simulate galaxy maps for the seventh and final version of the CFHTLS: the T0007\footnote{\texttt{http://www.cfht.hawaii.edu/Science/CFHLS/T0007/}} data release. The CFHTLS Wide consists of four independent fields covering, after masking, an effective area of 133$\sq^\circ$. After a magnitude cut at $i<22.5$, which ensures that our photometric redshift samples are 100$\%$ complete for all galaxy types and limit the number of outliers \citep{IlbertEtal06,CouponEtal09}, the entire sample contains 2,710,739 galaxies. We use a selection identical to \citet{WolkEtal13} by considering volume-limited samples with $M_{g}<-20.7$ in different redshift bins.
\newline
\indent
For the purposes of this paper, we restrict the analysis to the largest field, W1, in the photometric redshift bin $0.6 <z_{\textrm{phot}}< 0.8$, corresponding to the sample with the highest statistics. Thus this sample dominated the fit of the measured two-point correlation performed in \citet{WolkEtal13}, using the halo model \citep{SSHJ01, MaEtal00, PeacockEtal00, CoorayEtal02, CouponEtal12}, and the halo occupation distribution (HOD) parameters.
\newline
\indent
The W1 field is a square of $L = 7.46$ degrees on the side, and we divided it into $128^2$ square cells. This allows us to probe the galaxy angular power spectrum in the multipole range $60 \la \ell \la 4000$. 
For each cell, we determine in a Monte Carlo fashion the surface fraction,  that is the useful surface of the cell after masks due to bright stars, CCD faults, etc. are taken into account.
Cells with surface fraction less than $0.7$ were treated as completely masked, i.e. excluded from our analysis. The total number of galaxies in the map built in this way is 144,420. The map is shown on the left panel of Fig. \ref{Figmaps}.
 \subsection{Poisson sampling of a lognormal field} \label{Model}
\indent We model the galaxy count map as a Poisson sample of an underlying density field. The Poisson intensities are given in each cell by the integral over the cell of the field times the surface fraction of the cell not covered by the mask. The key assumption is then that the $128^2$-variate PDF describing the joint probability of the integrated field values behind the cells is a lognormal PDF. In other words, the map $\vecA = \ln \vecrho$ is a $128^2$ dimensional Gaussian vector. Its covariance matrix $\omega_A$ of size $128^2 \times 128^2$ follows the familiar relation for lognormal variables \citep[e.g.]{ColesJones91}
\beq \label{Lncov}
\omega_{A,ij} = \ln \lp1 + \omega_{ij} \rp,
\enq
where $\omega_{ij}$ is the covariance matrix of the dimensionless fluctuations of $\vecrho$. Further, $\omega_{ij}  = \omega(\vectheta_i-\vectheta_j)$ is the angular two-point function of the continuous density fluctuation field, filtered with a square top-hat filter of side the cell size. Note that due to the filtering the two-point-function is homogeneous but slightly anisotropic on scales similar to the cell size. A simulation of a galaxy map goes through the following steps.
\begin{enumerate}
\item We generate the $128^2$ field values $\rho_i = 1 + \delta_i$ from a lognormal PDF. The covariance matrix of the PDF is set to match precisely the best-fit HOD modeling of the (unfiltered) two-point function in \citet{WolkEtal13}. To generate $\vecrho$, a Gaussian vector $\vecA$ with covariance given in Eq. \eqref{Lncov} is generated in a $256^2$ map with the circulant embedding method, and the relevant sub-box exponentiated.  For the sake of completeness, Appendix \ref{filtered2pt} describes in  detail how the filtered two-point function entering Eq. \eqref{Lncov} was obtained from the unfiltered function.
\item We then generate counts in the $128^2$ cells following Poisson laws to create the galaxy map. The joint probability for the galaxy map $\vecN$ given $\vecrho$ is
\beq \label{sampling}
P(\vecN | \vecrho) =  \prod_{i }e^{-\bar N f_i \rho_i} \frac{\lp  \bar N f_i \rho_i\rp^{N_i}}{N_i!},
\enq
\end{enumerate}
where $f_i$ is the surface fraction of cell $i$ not covered by the mask. The Poisson intensity $\bar N$ is the ensemble mean number of galaxies in a cell with $f = 1$.
The right panel of Fig. \ref{Figmaps} shows one such simulation.
\subsection{Spectra} \label{Results}
\begin{figure}
\begin{center}
\includegraphics[width = 0.45\textwidth]{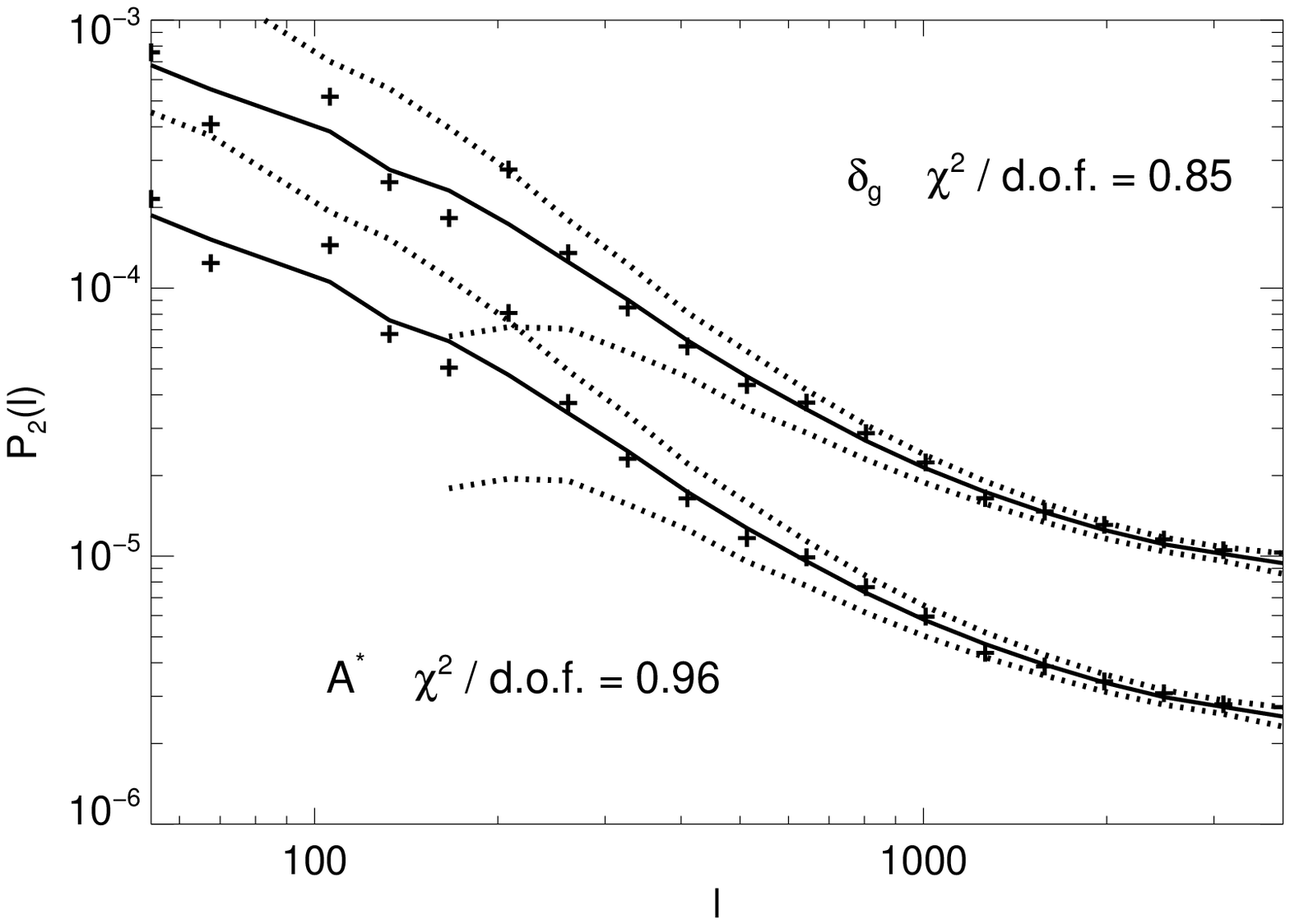}
\includegraphics[width = 0.45\textwidth]{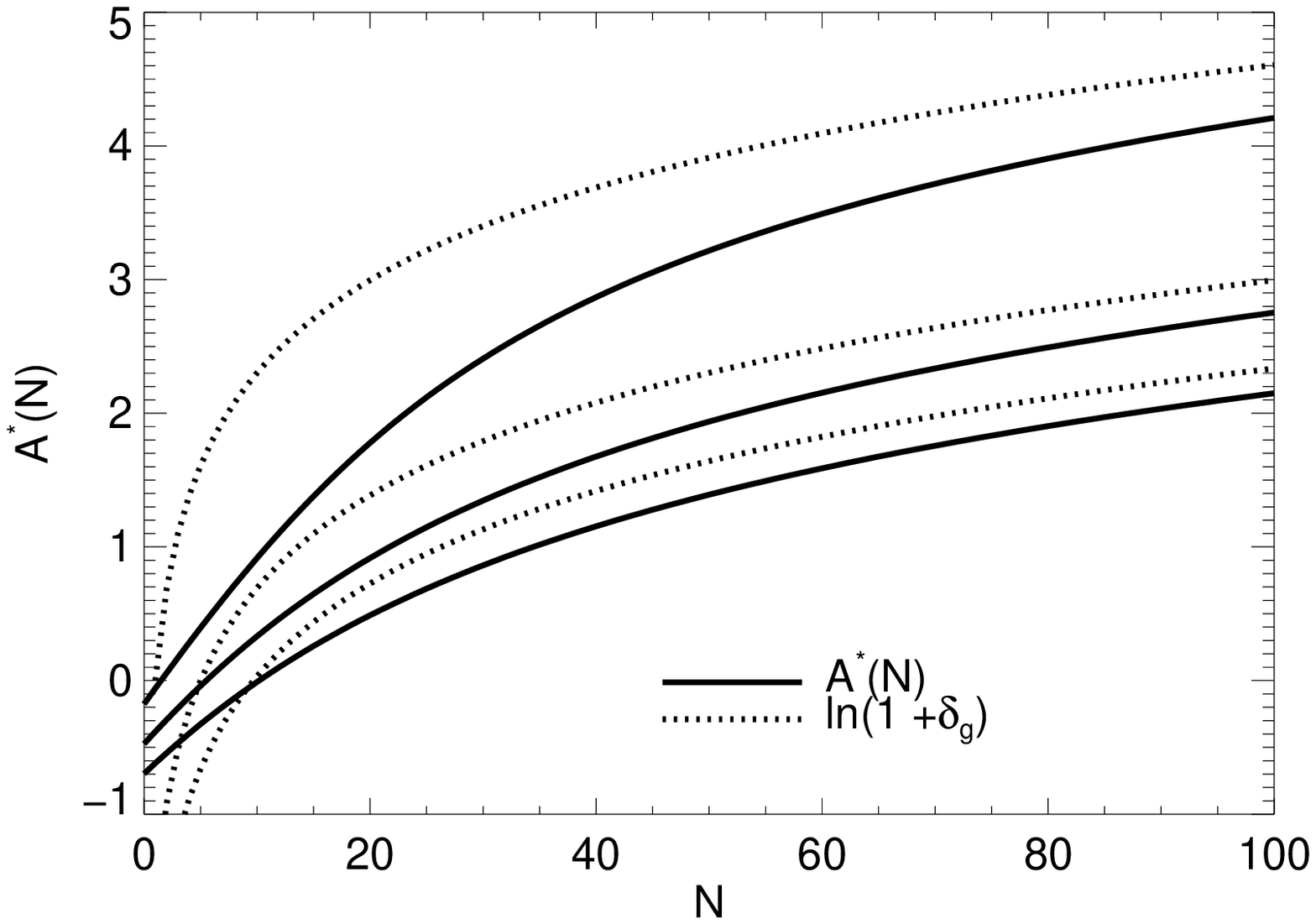}

\caption{Upper panel : The crosses in the upper set of lines show the angular power spectrum of the CFHTLS W1 field, including galaxies with photometric redshift estimate between 0.6 and 0.8 (see the map on the left panel of Fig. \ref{Figmaps}). The solid line shows the prediction of the Poisson-lognormal model. The covariance of the underlying lognormal variables is set to match the best-fit model to the two-point function performed by \citet{WolkEtal13}, no fit to the actual data points on this figure was performed. The $\chi^2$ per d.o.f. estimated with the model covariance matrix shows a consistent value of $0.85$. The dashed lines show the $\pm 2 \sigma$ region centered on the prediction. The lower set of lines on the upper panel show the same for the spectrum of the non-linearly transformed map $A^*(N)$ defined in section \ref{nlmap}. The lower panel shows as the solid lines $A^*(N)$ for three different values of $\bar N = 1,5$ and $9.65$, from top to bottom. The dashed lines show the logarithmic transform $\ln \lp 1 + \delta_g \rp$ (undefined for $N = 0$) for comparison.}
\label{Figpd}
\end{center}
\end{figure}

By construction the simulations have the two-point statistics originating from a fit by \citet{WolkEtal13}. A useful first test is whether the Fourier modes have the correct power. We proceed as follows. For a map $\boldsymbol \phi$, we calculate the power $\hat P^\phi_2(\ell)$ applying the discrete Fourier transform, squaring the magnitude of the result, and averaging over angle and multipoles,
\beq \label{Def spec}
\hat P^\phi_2(\ell) = \frac 1 V \frac{1}{N_\ell} \sum_{\ell' \in \Delta(\ell)} \left | \tilde \phi(\ell') \right|^2,
\enq
where $N_\ell$ is the number of modes in the sum. To evaluate the galaxy power spectrum, we evaluate Eq. \eqref{Def spec} for the map
\beq
 \delta_g = \frac{ N / f -\hat {\bar N} }{\hat {\bar N}},\quad f \ne 0,
\enq
and $\delta_{g}$ is set to zero wherever $f$  is zero. In this expression, the Poisson intensity $\bar N$ is estimated with
\beq \label{Estim}
 \hat{\bar N} = \frac{N_{\textrm{tot}}}{\sum_i{f_i}},
\enq
where $N_{\textrm{tot}}$ is the total number of galaxies in the map under consideration. Application to the data map gives $\bar N = 9.65$.
We do not subtract any shot noise term, playing no role for the purposes of this paper. We use 20 $\ell$-bins equally spaced in $\ln \ell$. We then evaluate the predictions for the spectrum from the Poisson-lognormal model, together with its covariance matrix directly from a very large number of simulations. The upper set of lines on the upper panel of Fig. \eqref{Figpd} show the predictions and measurement from the data. The crosses indicate the measurement, and the predictions are shown as the upper solid line, together with the  $2\sigma$ confidence region (dashed). A $\chi^2$ value per d.o.f. of  $0.85$ is obtained with the full covariance. While the overall agreement of the spectra is not that surprising, the value of the $\chi^2$ indicates that the model captures well the covariance of the measurements.

\subsection{Comparison with $k$-space sampling method}
\begin{figure}
\begin{center}
\includegraphics[width = 0.45\textwidth]{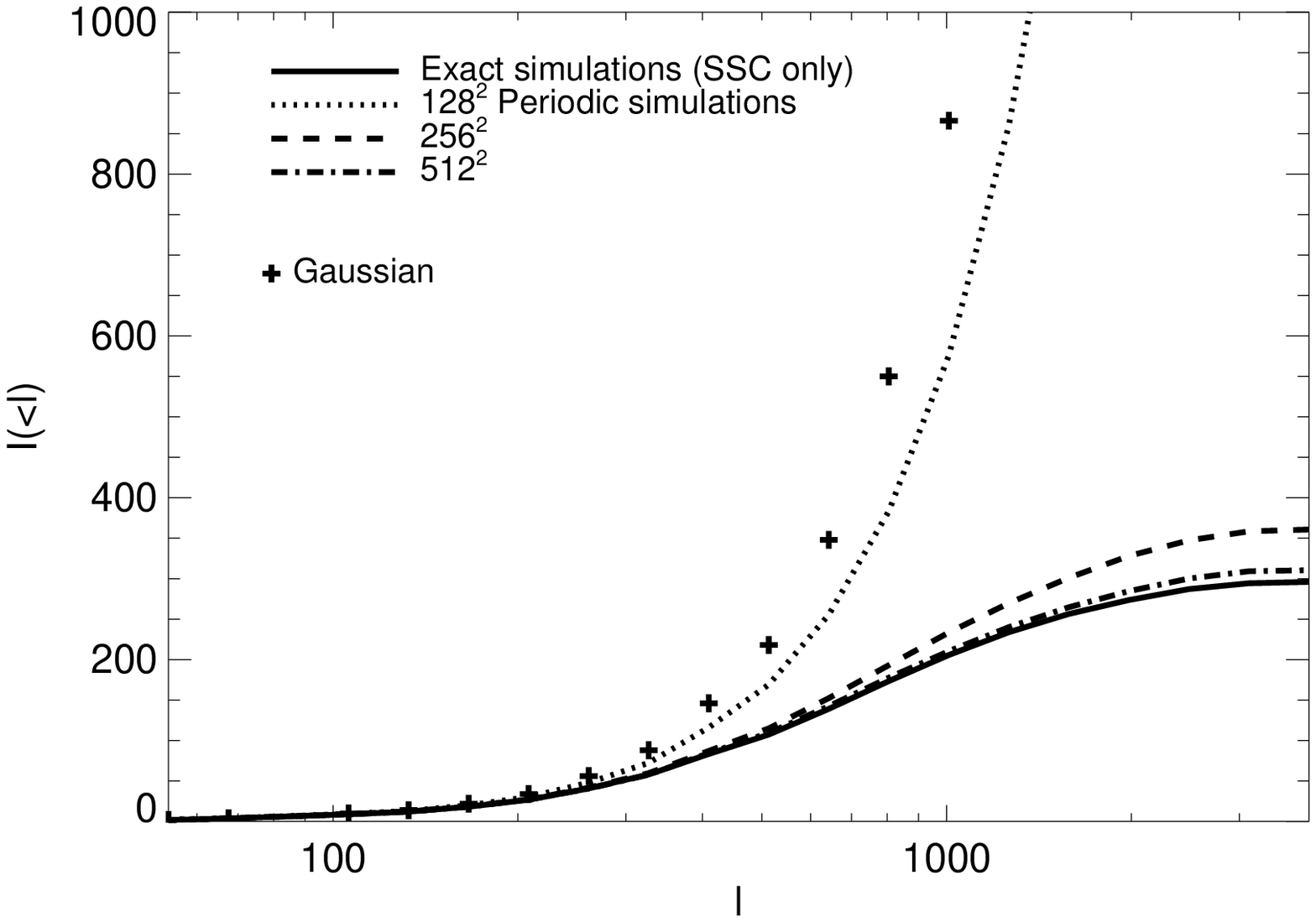}
\includegraphics[width = 0.45\textwidth]{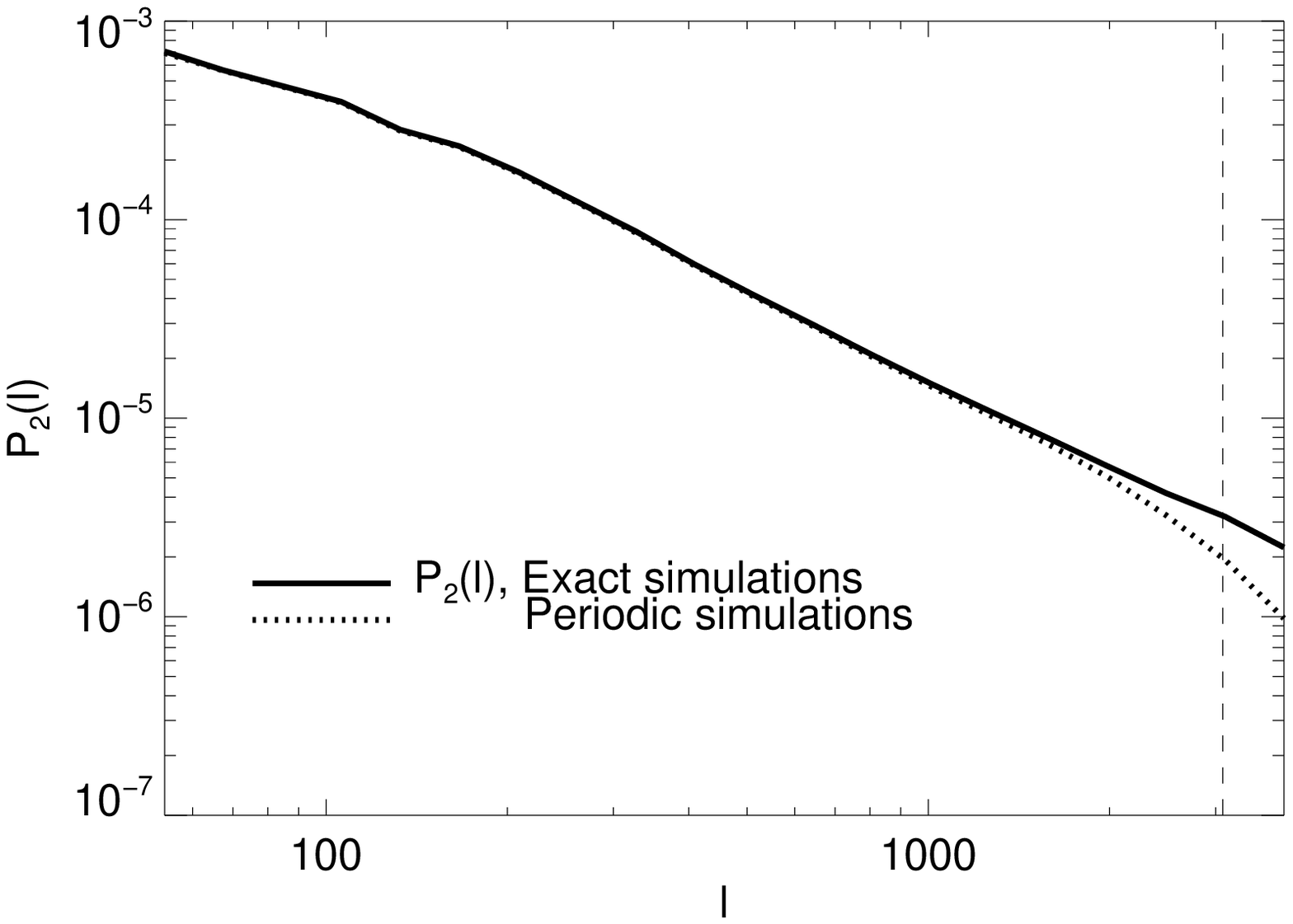}

\caption{Comparison of periodic against exact simulations achieved with the methods of this paper. Upper panel : the solid line shows the cumulative information content of the spectrum of the underlying density field in our lognormal model. The crosses indicate the Gaussian regime. The dotted, dashed and dotted-dashed lines were generated using $k$- space sampling, embedding the map in a larger periodic box as indicated. These lack box to box fluctuations. These curves show the impact of beat-coupling, or super-sample covariance on the spectrum covariance matrix. The local average effect  \citep{DePutterEtal12} is not included there, its inclusion would change the value of the plateau. Lower panel : the spectrum of the underlying density fluctuation map, using the exact circulant embedding method (solid) and the $k$-space sampling method (dotted). The latter lacks close to the pixel scale $\ell \sim \pi N /L $, shown as the vertical line, the additional power due to the grid discreteness.\label{Figper}}
\end{center}
\end{figure}

An interesting quantity related to the covariance matrix of the matter power spectrum is the squared signal to noise $\lp S/N\rp^2$, or cumulative information \citep{RimesHamilton05,RimesHamilton06,LeePen08}
\beq \label{cuminfo}
I(\le \ell) = \sum_{\ell_1,\ell_2 \le\ell} \lb \vecC^{-1} \rb_{\ell_1,\ell_2},
\enq
where 
\beq
\vecC_{\ell_1\ell_2} =\frac{\av{\hat P_2(\ell_1) \hat P_2(\ell_2)} }{ \av{\hat P_2(\ell_1)}\av{\hat P_2{(\ell_2)}}}-1,\quad \ell_1,\ell_2 \le \ell.
\enq
$I(\le \ell)$ can be interpreted as the Fisher information content of the spectrum on a log-amplitude parameter. For Gaussian statistics $I(\le \ell)$ is half the number of modes below $\ell$. As discussed in the introduction, super-survey modes cause a strong saturation of the information. These modes are not carefully reproduced by more traditional '$k$-space sampling' methods. To investigate this, we simulate $\vecA$ on a periodic box where the Fourier modes are set to match those of the continuous field, 
\beq \label{Periodicboxmodes}
P^A_2(\vecl) = \int_{-\infty}^{\infty} \int_{-\infty}^{\infty} d^2\theta\: \ln (1 + \omega(\vectheta)) e^{-i\vecl\cdot \vectheta}.
\enq 
We embed the $128^2$ box of interest in a larger periodic box and extract the spectra of $\vecrho$ according to Eq. \eqref{Def spec}. The $\vecl = 0$ mode (the 'DC mode' \citep{Sirko05,GnedinEtal11}) of the larger box is set to zero, effectively forcing all simulation to have identical mean density. The dotted, dashed and dot-dashed lines in the upper panel of Fig. \ref{Figper} show from top to bottom the cumulative information in the $\vecrho$ map when using a $128^2$, $256^2$ and $512^2$ periodic box to simulate the map. The curves converge slowly to the exact result, demonstrating clearly the importance of implementing the super-surveys modes for the spectrum covariance. As a technical comment, it might be useful to note that in that figure, in the nomenclature of \cite{DePutterEtal12} only the beat-coupling effect, or super-sample covariance is included, but not the local average (also known as the integral constraint) effect. The curves on the lower panel show the spectrum of $\rho$ with the circulant embedding method (solid), and with the $k$-sampling method (dashed). The latter lacks the power due to the grid discreteness on scales close to the pixel scale $N\pi/L$, indicated by the vertical line. This could only be corrected with $k$-space sampling by simulating the same volume at a higher resolution, at further additional cost. Also, note that the input spectrum in Eq. \eqref{Periodicboxmodes} differs from that of the circulant embedding method through (i) an integration rather than a discrete FFT (ii) the inclusion of the entire space rather than $[-L,L]^2$. Thus, the circulant embedding method is not only exact, but in fact also cheaper and more straightforward to implement .
\subsection{Consistency of Poisson-lognormal statistics beyond the power spectrum}
The spectrum and its covariance matrix test the data up to the four point statistics. We now show that the model in fact captures much more statistical properties correctly. We will consider the count in cells (CIC) of the map, as well as the local non-linear transform $A^*(N)$ introduced in \cite{CarronSzapudi14}, that was designed to recapture most parameter information in the data in its mean and spectrum. In that sense, $A^*$ is the analog of the log-transform $A = \ln (1 + \delta)$ as in \cite{Neyrinck11} of the matter fluctuation field in the presence of Poissonian discreteness noise.
\subsubsection*{Count in cells}
\begin{figure}
\begin{center}
\includegraphics[width = 0.45\textwidth]{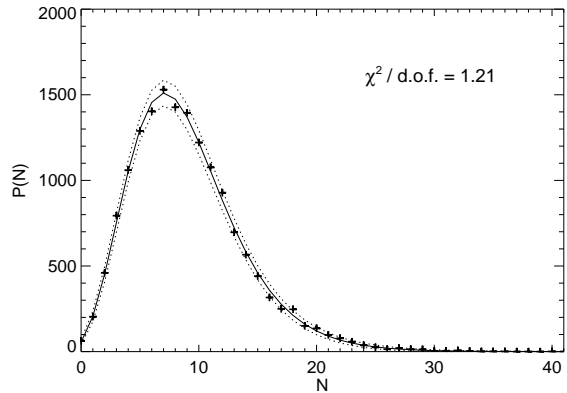}
\caption{
The crosses show the histogram of the counts in cell in the  CFHTLS W1 field, including galaxies with photometric redshift estimate between 0.6 and 0.8 (see the map on the left panel of Fig. \ref{Figmaps}). The black line shows the prediction for the histogram for a Poisson-sampling of a lognormal field, obtained together with its covariance matrix by averaging over a large number of data simulations (such as the right panel of Fig. \ref{Figmaps}). The covariance matrix of the underlying lognormal variables (the only free parameters of the model) is set to match the best-fit model to the two-point function performed by \citet{WolkEtal13}, no fit to the data points on this figure was performed. The dashed lines show the $\pm 2 \sigma$ region centered on the prediction.
\label{figcic}}

\end{center}
\end{figure}

We measure the histogram
\beq
\hat P(N) = \sum_{f_i \ne 0} \delta_{N_iN}
\enq
of the map and of a very large  number of simulations, together with its covariance matrix. Cells that are completely masked are ignored. Following Eq. \eqref{sampling} and the lognormal field assumption, the mean value of the histogram  is given by a mixture of Poisson-lognormal distributions with rates $\bar \vecN= \bar N \vecf$,
\beq
\begin{split}
\av{\hat P(N)} &= \int_{-\infty}^{\infty}\frac{dA}{\sqrt{2\pi \sigma^2_A}} e^{-\frac1{2\sigma^2_A} \lp  A + \frac 12 \sigma_A^2\rp^2 } \\ 
&\quad  \frac{e^{NA}}{N!}\cdot  \sum_{f_i \ne 0} \bar N_i^N\:e^{- \bar N_i e^A }
\end{split}
\enq
We use $N$ from $0$ to $41$, which is the range of the data map on Fig. \ref{Figmaps}. The results are shown on Fig. \ref{figcic}. The crosses show the data, the solid lines the model predictions and the dashed lines the $2\sigma$ errors centered on the predictions. Again, the $\chi^2$ per d.o.f., $1.21$, shows a perfectly consistent value.
\subsubsection*{Non-linear transform $A^*$ }\label{nlmap}
The transform $A^*(N)$ is defined as the solution of the non-linear equation
\beq \label{As}
e^{A^*}\:\bar N \sigma^2_A  + A^* =\sigma^2_A\lp N - 1/2 \rp.
\enq
For large values of $N$, $A^*$ becomes the logarithmic mapping $\ln(1 + \delta_g)$, and corrects the inadequacy of the latter for small values of $N$, by taking into account the noise properties to recover as most information as possible, see \cite{CarronSzapudi14} for details. It can also be interpreted as the univariate version of the maximum a posteriori solution for the reconstruction of the underlying density field, see \cite{KitauraEtal10}.
\newline
\indent
We transform in each cell $N$ to $A^*$, according to the above equation. In that equation, the parameter $\sigma^2_A$ is identical for each cell, and, according to Eq. \eqref{Lncov}, is equal to $\ln(1 + \sigma^2_\delta) \sim 0.13$. On the other hand, the Poisson intensities vary from cell to cell due to the masking, according to $\bar N_i = \bar N f_i$, where $\bar N \sim 9.68$. The lower panel of Fig. \ref{Figpd} shows as the solid lines the $A^*(N)$ mapping for $\bar N = 1,5$ and $9.68$, from top to bottom. The dashed line show the mapping $\ln (1 + \delta_g) = \ln \lp N /\bar N \rp$ for comparison. Note that the latter is not well defined for $N = 0$. We then extract the power spectrum of $A^*$, shown in the lower set of curves in the upper panel of the same figure. Again, the dashed lines show $2 \sigma$ deviations. The spectra of $\delta_g$ and $A^*$ are very similar on large scales, as expected, and deviates only slightly at the smallest scales. The $\chi^2$ per d.o.f. becomes $0.96$.

 \section{Summary and discussion} \label{Conclusions}
 We have presented an exact method to simulate Gaussian variables on a regular grid in a finite volume. 
 Based on embedding the volume in a periodic volume of twice the side with an adequately modified power spectrum, it is fast yet implements exactly the statistical properties of the variables on the grid points. In particular, box-to-box mean fluctuations and super-survey modes are automatically present in the simulations at no additional cost. A minimal modification with respect to codes using periodic grid simulations is required.  The method is limited to the generation of Gaussian fields or simple transformations thereof, and cannot reproduce the finest properties of cosmological fields. Nevertheless, it it one ideal tool to produce fast realizations of simplified statistical models  using Gaussian fields as an ingredient. A typical application of the technique is the estimation of cosmic variance and covariance.
 \newline
\indent In this paper, we have applied the technique to produce simulations of the CFTHLS W1 projected count maps, using as model the Poisson sampling of a lognormal field. We measured the spectrum, its covariance, the counts in cells PDF and the spectrum of the $A^*$ non-linear transform. All showed perfect consistency between the data and our simulations. While the Poisson sampling of a lognormal field is approximate for three dimensional surveys \citep[e.g.]{WildEtal05,KitauraEtal10}, it appears to be providing a surprisingly precise model for projected data. In addition, we demonstrated how important the super-surveys modes are to capture correctly the 'beat-coupling' contribution to the covariance of the spectrum, reducing the independent information content of non-linear scales. These modes are absent, or attenuated, using traditional $k$-space sampling methods. We note that in general  both the choice of a Poisson-lognormal model and the $A^*$ non-linear transform is tied to the angular size of the cells. The question whether the Poisson-lognormal field can still describe consistently the behavior of the transform in the data for other cell sizes is left for future investigations.
\newline
\indent Among further applications of the method, it would be interesting to see how well the 3-dimensional version of the model can reproduce galaxy surveys data and super-survey modes with the circulant embedding method. Such lognormal mock catalogues are already used for the estimation of covariance matrices, although with periodic grids  \citep{ColeEtal05,BlakeEtal11,BeutlerEtal11}. Also, since the model is a great match to the data, it can be used to test the covariance and information content other statistics, such as higher order moments, logarithmic transforms \citep{NeyrinckEtal09}, the information-sufficient statistics introduced in \cite{CarronSzapudi13,CarronSzapudi14}, to constrain HOD and/or cosmological parameters. This will be reported in a forthcoming publication \citep{WolkEtal14}.  Of course, the technique can also readily be applied to simulate in a completely analogous manner other probes of cosmology where fast simulations of the same type are useful, such as weak lensing \citep{SimonEtal04,TakahashiEtal14}, the Lyman-$\alpha$ forest \citep{FontRiberaEtal12}, 21cm intensity mapping \citep{AlonsoEtal14} or patches of the CMB sky. 
\section*{Acknowledgments}
The authors acknowledge NASA grants NNX12AF83G and NNX10AD53G for support.
\newline
\indent
Part of this work was based on observations obtained with
MegaPrime/MegaCam, a joint project of CFHT and CEA/IRFU, at the
Canada-France-Hawaii Telescope (CFHT) which is operated by the
National Research Council (NRC) of Canada, the Institut National des
Science de l'Univers of the Centre National de la Recherche
Scientifique (CNRS) of France, and the University of Hawaii. This work
is based in part on data products produced at Terapix available at the
Canadian Astronomy Data Centre as part of the Canada-France-Hawaii
Telescope Legacy Survey, a collaborative project of NRC and CNRS.

\appendix
\section{Two-point function of the filtered density field} \label{filtered2pt}
\newcommand{\vecphi}{\boldsymbol \phi}
\newcommand{\vecpsi}{\boldsymbol \psi}

In this section we describe how we obtain in an accurate yet fast way the anisotropic two-point function $\omega^\ell(\vectheta)$ of the square top-hat of side $\ell$ filtered density field from its unfiltered isotropic counterpart $\omega(\theta)$.
\newline
\indent
A filtered two-point function is given by
\beq
\omega^\ell(\vectheta) = \int d^2\phi\:W(\vecphi)\int d^2\psi\:W(\vecpsi) \omega(\vectheta + \vecphi-\vecpsi),
\enq
where in our case the filter is
\beq
W(\vectheta)  = \frac 1 {\ell^2} \begin{cases} 1, \textrm{  if  } -\frac \ell 2 \le \theta_1,\theta_2 \le \frac \ell 2 \\ 0,\textrm{  if not}\end{cases}.
\enq
After performing the overlap integral of the filter function, this becomes
\beq \label{exactformula}
\omega^\ell(\vectheta)  = \int_{[-1,1]^2} d^2\epsilon \lp 1 - |\epsilon_1|\rp \lp 1 - |\epsilon_2|\rp \:\omega\lp|\vectheta + \ell \vecepsilon |\rp.
\enq
The special case $\vectheta = 0$ can  be reduced further
\beq \label{xibar}
\omega^\ell(0) = 4 \int_0^{\sqrt{2}} dx \: x\: \omega(\ell x)f(x),
\enq
with
\beq
f(x) = \begin{cases} \frac \pi 2 +\frac 12x^2 -2x, \quad  x\le 1 \\ \frac \pi 2 -\frac 12 x^2 -2 \textrm{acos}\lp \frac 1 x \rp + 2\sqrt{x^2-1} -1,\quad x \ge 1. \end{cases}
\enq
For distances comparable to $\ell$ we used direct numerical integration of formulae \eqref{exactformula} and \eqref{xibar} with an accurate Gauss-Legendre integration scheme. Note that at these distances $\omega_\delta^\ell$ is a function of both magnitude and angle. At larger distances $(\theta/\ell \sim 15)$ we used the derivative expansion (in analogy to \citet{VogeleySzalay96})
\beq \label{CoolEq}
\omega^\ell(\vectheta)\sim \omega(\theta) + \frac{1}{12} \lp \frac \ell\theta \rp^2 \frac{\partial^2 \omega(\theta)}{\partial \ln \theta^2 } + O\lp\frac{\ell}{\theta}\rp^4.
\enq
The second logarithmic derivative was obtained using $O(h^4)$ 4-points finite differences on a equidistant logarithmic grid. For a perfect power-law $\omega(\theta) \propto \theta^{1-\gamma}$, this derivative is simply $\omega(\theta)$ times the power-law exponent squared. We obtain that the correction with respect to the unfiltered two-point function is
\beq
\frac{\omega^\ell(\vectheta)}{\omega(\theta)} -1 \sim \frac{1}{12} \lp \frac \ell \theta \rp^2 (1-\gamma)^2.
\enq
In our case $\gamma \sim 1.7$ \citep{WolkEtal13}. This shows that the first correction in \ref{CoolEq} is already less than one part in one thousand for $\theta = 10\:\ell$.

\bibliographystyle{mn2e}
\bibliography{bib}

\begin{thebibliography}{}

\bibitem[\protect\citeauthoryear{{Alonso}, {Ferreira} \& {Santos}}{{Alonso}
  et~al.}{2014}]{AlonsoEtal14}
{Alonso} D.,  {Ferreira} P.~G.,    {Santos} M.~G.,  2014, ArXiv e-prints
  astro-ph/1405.1751

\bibitem[\protect\citeauthoryear{{Bernardeau} \& {Kofman}}{{Bernardeau} \&
  {Kofman}}{1995}]{BernardeauKofman95}
{Bernardeau} F.,  {Kofman} L.,  1995, \apj, 443, 479

\bibitem[\protect\citeauthoryear{{Bertschinger}}{{Bertschinger}}{2001}]{Bertsc%
hinger01}
{Bertschinger} E.,  2001, \apjs, 137, 1

\bibitem[\protect\citeauthoryear{{Beutler} \& et al.}{{Beutler} \&
  et~al.}{2011}]{BeutlerEtal11}
{Beutler} F.,  et al. 2011, \mnras, 416, 3017

\bibitem[\protect\citeauthoryear{{Blake} \& et al}{{Blake} \&
  et~al}{2011}]{BlakeEtal11}
{Blake} C.,  et al 2011, \mnras, 415, 2892

\bibitem[\protect\citeauthoryear{{Carron}}{{Carron}}{2011}]{Carron11}
{Carron} J.,  2011, \apj, 738, 86

\bibitem[\protect\citeauthoryear{{Carron} \& {Neyrinck}}{{Carron} \&
  {Neyrinck}}{2012}]{CarronNeyrinck12}
{Carron} J.,  {Neyrinck} M.~C.,  2012, \apj, 750, 28

\bibitem[\protect\citeauthoryear{{Carron} \& {Szapudi}}{{Carron} \&
  {Szapudi}}{2013}]{CarronSzapudi13}
{Carron} J.,  {Szapudi} I.,  2013, \mnras, 434, 2961

\bibitem[\protect\citeauthoryear{{Carron} \& {Szapudi}}{{Carron} \&
  {Szapudi}}{2014}]{CarronSzapudi14}
{Carron} J.,  {Szapudi} I.,  2014, \mnras, 439, L11

\bibitem[\protect\citeauthoryear{{Cole} \& et al.}{{Cole} \&
  et~al.}{2005}]{ColeEtal05}
{Cole} S.,  et al. 2005, \mnras, 362, 505

\bibitem[\protect\citeauthoryear{{Coles} \& {Jones}}{{Coles} \&
  {Jones}}{1991}]{ColesJones91}
{Coles} P.,  {Jones} B.,  1991, \mnras, 248, 1

\bibitem[\protect\citeauthoryear{{Cooray} \& {Sheth}}{{Cooray} \&
  {Sheth}}{2002}]{CoorayEtal02}
{Cooray} A.,  {Sheth} R.,  2002, Phys. Rep., 372, 1

\bibitem[\protect\citeauthoryear{{Coupon}, {Kilbinger}, {McCracken}, {Ilbert},
  {Arnouts}, {Mellier}, {Abbas}, {de la Torre}, {Goranova}, {Hudelot}, {Kneib}
  \& {Le F{\`e}vre}}{{Coupon} et~al.}{2012}]{CouponEtal12}
{Coupon} J.,  {Kilbinger} M.,  {McCracken} H.~J.,  {Ilbert} O.,  {Arnouts} S.,
  {Mellier} Y.,  {Abbas} U.,  {de la Torre} S.,  {Goranova} Y.,  {Hudelot} P.,
  {Kneib} J.-P.,    {Le F{\`e}vre} O.,  2012, \aap, 542, A5

\bibitem[\protect\citeauthoryear{{Coupon}}{{Coupon}}{2009}]{CouponEtal09}
{Coupon} J. e.~a.,  2009, \aap, 500, 981

\bibitem[\protect\citeauthoryear{{de Putter}, {Wagner}, {Mena}, {Verde} \&
  {Percival}}{{de Putter} et~al.}{2012}]{DePutterEtal12}
{de Putter} R.,  {Wagner} C.,  {Mena} O.,  {Verde} L.,    {Percival} W.~J.,
  2012, \jcap, 4, 19

\bibitem[\protect\citeauthoryear{{Dietrich} \& {Newsam}}{{Dietrich} \&
  {Newsam}}{1993}]{DietrichNewsam93}
{Dietrich} C.~R.,  {Newsam} G.~N.,  1993, Water Resources Research, 29, 2861

\bibitem[\protect\citeauthoryear{Dietrich \& Newsam}{Dietrich \&
  Newsam}{1997}]{DietrichNewsam97}
Dietrich C.~R.,  Newsam G.~N.,  1997, SIAM J. Sci. Comput., 18, 1088

\bibitem[\protect\citeauthoryear{{Font-Ribera}, {McDonald} \&
  {Miralda-Escud{\'e}}}{{Font-Ribera} et~al.}{2012}]{FontRiberaEtal12}
{Font-Ribera} A.,  {McDonald} P.,    {Miralda-Escud{\'e}} J.,  2012, \jcap, 1,
  1

\bibitem[\protect\citeauthoryear{{Gnedin}, {Kravtsov} \& {Rudd}}{{Gnedin}
  et~al.}{2011}]{GnedinEtal11}
{Gnedin} N.~Y.,  {Kravtsov} A.~V.,    {Rudd} D.~H.,  2011, \apjs, 194, 46

\bibitem[\protect\citeauthoryear{{Hamilton}, {Rimes} \&
  {Scoccimarro}}{{Hamilton} et~al.}{2006}]{HamiltonEtal06}
{Hamilton} A.~J.~S.,  {Rimes} C.~D.,    {Scoccimarro} R.,  2006, \mnras, 371,
  1188

\bibitem[\protect\citeauthoryear{{Ilbert}}{{Ilbert}}{2006}]{IlbertEtal06}
{Ilbert} O. e.~a.,  2006, \aap, 457, 841

\bibitem[\protect\citeauthoryear{{Kayo}, {Taruya} \& {Suto}}{{Kayo}
  et~al.}{2001}]{KayoEtal01}
{Kayo} I.,  {Taruya} A.,    {Suto} Y.,  2001, \apj, 561, 22

\bibitem[\protect\citeauthoryear{{Kitaura}, {Jasche} \& {Metcalf}}{{Kitaura}
  et~al.}{2010}]{KitauraEtal10}
{Kitaura} F.-S.,  {Jasche} J.,    {Metcalf} R.~B.,  2010, \mnras, 403, 589

\bibitem[\protect\citeauthoryear{{Kofman}, {Bertschinger}, {Gelb}, {Nusser} \&
  {Dekel}}{{Kofman} et~al.}{1994}]{KofmanEtal94}
{Kofman} L.,  {Bertschinger} E.,  {Gelb} J.~M.,  {Nusser} A.,    {Dekel} A.,
  1994, \apj, 420, 44

\bibitem[\protect\citeauthoryear{{Lee} \& {Pen}}{{Lee} \&
  {Pen}}{2008}]{LeePen08}
{Lee} J.,  {Pen} U.,  2008, \apjl, 686, L1

\bibitem[\protect\citeauthoryear{{Li}, {Hu} \& {Takada}}{{Li}
  et~al.}{2014}]{LiEtal14}
{Li} Y.,  {Hu} W.,    {Takada} M.,  2014, \prd, 89, 083519

\bibitem[\protect\citeauthoryear{{Ma} \& {Fry}}{{Ma} \& {Fry}}{2000}]{MaEtal00}
{Ma} C.-P.,  {Fry} J.~N.,  2000, \apj, 543, 503

\bibitem[\protect\citeauthoryear{{Neyrinck}}{{Neyrinck}}{2011}]{Neyrinck11}
{Neyrinck} M.~C.,  2011, \apj, 742, 91

\bibitem[\protect\citeauthoryear{{Neyrinck}, {Szapudi} \& {Rimes}}{{Neyrinck}
  et~al.}{2006}]{NeyrinckEtal06}
{Neyrinck} M.~C.,  {Szapudi} I.,    {Rimes} C.~D.,  2006, \mnras, 370, L66

\bibitem[\protect\citeauthoryear{{Neyrinck}, {Szapudi} \& {Szalay}}{{Neyrinck}
  et~al.}{2009}]{NeyrinckEtal09}
{Neyrinck} M.~C.,  {Szapudi} I.,    {Szalay} A.~S.,  2009, \apjl, 698, L90

\bibitem[\protect\citeauthoryear{{Peacock} \& {Smith}}{{Peacock} \&
  {Smith}}{2000}]{PeacockEtal00}
{Peacock} J.~A.,  {Smith} R.~E.,  2000, \mnras, 318, 1144

\bibitem[\protect\citeauthoryear{{Pen}}{{Pen}}{1997}]{Pen97}
{Pen} U.-L.,  1997, \apjl, 490, L127

\bibitem[\protect\citeauthoryear{Press, Teukolsky, Vetterling \&
  Flannery}{Press et~al.}{2007}]{PressEtal07}
Press W.~H.,  Teukolsky S.~A.,  Vetterling W.~T.,    Flannery B.~P.,  2007,
  Numerical Recipes 3rd Edition: The Art of Scientific Computing.
Cambridge University Press, New York, NY, USA

\bibitem[\protect\citeauthoryear{{Rimes} \& {Hamilton}}{{Rimes} \&
  {Hamilton}}{2005}]{RimesHamilton05}
{Rimes} C.~D.,  {Hamilton} A.~J.~S.,  2005, \mnras, 360, L82

\bibitem[\protect\citeauthoryear{{Rimes} \& {Hamilton}}{{Rimes} \&
  {Hamilton}}{2006}]{RimesHamilton06}
{Rimes} C.~D.,  {Hamilton} A.~J.~S.,  2006, \mnras, 371, 1205

\bibitem[\protect\citeauthoryear{{Schneider}, {Cole}, {Frenk} \&
  {Szapudi}}{{Schneider} et~al.}{2011}]{SchneiderEtal11}
{Schneider} M.~D.,  {Cole} S.,  {Frenk} C.~S.,    {Szapudi} I.,  2011, \apj,
  737, 11

\bibitem[\protect\citeauthoryear{{Scoccimarro}, {Sheth}, {Hui} \&
  {Jain}}{{Scoccimarro} et~al.}{2001}]{SSHJ01}
{Scoccimarro} R.,  {Sheth} R.~K.,  {Hui} L.,    {Jain} B.,  2001, ApJ, 546, 20

\bibitem[\protect\citeauthoryear{{Sefusatti}, {Crocce}, {Pueblas} \&
  {Scoccimarro}}{{Sefusatti} et~al.}{2006}]{SefusattiEtal06}
{Sefusatti} E.,  {Crocce} M.,  {Pueblas} S.,    {Scoccimarro} R.,  2006, \prd,
  74, 023522

\bibitem[\protect\citeauthoryear{{Simon}, {King} \& {Schneider}}{{Simon}
  et~al.}{2004}]{SimonEtal04}
{Simon} P.,  {King} L.~J.,    {Schneider} P.,  2004, \aap, 417, 873

\bibitem[\protect\citeauthoryear{{Sirko}}{{Sirko}}{2005}]{Sirko05}
{Sirko} E.,  2005, \apj, 634, 728

\bibitem[\protect\citeauthoryear{{Takada} \& {Hu}}{{Takada} \&
  {Hu}}{2013}]{TakadaHu13}
{Takada} M.,  {Hu} W.,  2013, \prd, 87, 123504

\bibitem[\protect\citeauthoryear{{Takahashi}, {Soma}, {Takada} \&
  {Kayo}}{{Takahashi} et~al.}{2014}]{TakahashiEtal14}
{Takahashi} R.,  {Soma} S.,  {Takada} M.,    {Kayo} I.,  2014, ArXiv e-prints
  astro-ph / 1405.2666

\bibitem[\protect\citeauthoryear{{Takahashi}, {Yoshida}, {Takada}, {Matsubara},
  {Sugiyama}, {Kayo}, {Nishizawa}, {Nishimichi}, {Saito} \&
  {Taruya}}{{Takahashi} et~al.}{2009}]{TakahashiEtal09}
{Takahashi} R.,  {Yoshida} N.,  {Takada} M.,  {Matsubara} T.,  {Sugiyama} N.,
  {Kayo} I.,  {Nishizawa} A.~J.,  {Nishimichi} T.,  {Saito} S.,    {Taruya} A.,
   2009, \apj, 700, 479

\bibitem[\protect\citeauthoryear{{Vogeley} \& {Szalay}}{{Vogeley} \&
  {Szalay}}{1996}]{VogeleySzalay96}
{Vogeley} M.~S.,  {Szalay} A.~S.,  1996, \apj, 465, 34

\bibitem[\protect\citeauthoryear{{Wild}, {Peacock}, {Lahav}, {Conway},
  {Maddox}, {Baldry}, {Baugh} \& other authors}{{Wild}
  et~al.}{2005}]{WildEtal05}
{Wild} V.,  {Peacock} J.~A.,  {Lahav} O.,  {Conway} E.,  {Maddox} S.,  {Baldry}
  I.~K.,  {Baugh}   other authors 2005, \mnras, 356, 247

\bibitem[\protect\citeauthoryear{Wolk \& et al.}{Wolk \&
  et~al.}{2014}]{WolkEtal14}
Wolk M.,  et al. 2014, in preparation

\bibitem[\protect\citeauthoryear{{Wolk}, {McCracken}, {Colombi}, {Fry},
  {Kilbinger}, {Hudelot}, {Mellier} \& {Ilbert}}{{Wolk}
  et~al.}{2013}]{WolkEtal13}
{Wolk} M.,  {McCracken} H.~J.,  {Colombi} S.,  {Fry} J.~N.,  {Kilbinger} M.,
  {Hudelot} P.,  {Mellier} Y.,    {Ilbert} O.,  2013, \mnras, 435, 2

\bibitem[\protect\citeauthoryear{Wood \& Chan}{Wood \& Chan}{1994}]{WoodChan94}
Wood A. T.~A.,  Chan G.,  1994, Journal of Computational and Graphical
  Statistics, 3, 409

\end{thebibliography}

\end{document}